\documentstyle[12pt,epsfig,axodraw]{article}
\renewcommand{\baselinestretch}{2.0}
\begin{document}
\newcommand {\be}{\begin{equation}}
\newcommand {\ee}{\end{equation}}
\newcommand {\ba}{\begin{eqnarray}}
\newcommand {\ea}{\end{eqnarray}}
\newcommand {\bea}{\begin{array}}
\newcommand {\cl}{\centerline}
\newcommand {\pl} {Phys.~Lett.} 
\newcommand {\prl} {Phys.~Rev.~Lett.}
\newcommand {\pr} {Phys.~Rev.}
\newcommand {\np} {Nucl.~Phys.}
\newcommand {\eea}{\end{array}}
\renewcommand {\thefootnote}{\fnsymbol{footnote}}

\vskip .5cm

\renewcommand 
{\thefootnote}{\fnsymbol{footnote}}
\def \a'{\alpha'}
\baselineskip 0.65 cm

\begin{flushright}
SISSA/86/2003/EP
\\
SLAC-PUB-10189 \\
hep-ph/0310055 \\
\today
\end{flushright} 
\begin{center}
{\Large{\bf
 Effects of the  Neutrino $B$-term on  Slepton 
Mixing 
and  
Electric Dipole Moments\footnote{Work supported by 
the U. S. 
Department
of Energy under
contract DE-AC03-76SF00515.}}}
{\vskip 0.5 cm}

{\bf {Yasaman  Farzan\\}}
{\vskip 0.5 cm}
{
{\it Stanford Linear Accelerator Center,
                                              2575 Sand 
Hill Road,
                                             Menlo Park, 
CA 94025
 }}
{\vskip 0.5 cm}
and \\
{  {\it Scuola
Internazionale superiore di Studi Avanzati,  via Beirut 
4, I-34014
Trieste, Italy}}
\vskip 0.5 cm
{\tt email:yasaman@slac.stanford.edu}
\end{center}
\renewcommand{\baselinestretch}{1.0}
\begin{abstract}
The supersymmetric standard model with 
right-handed
neutrino supermultiplets
generically contains a soft supersymmetry 
breaking mass term: $\delta L = B_\nu 
M\tilde{\nu}_R \tilde {\nu}_R/2$. We call this operator
the ``neutrino $B$-term".
We show that the neutrino $B$-term can give the 
dominant 
effects from the neutrino sector to 
lepton-flavor-violating processes and to lepton 
electric dipole moments.
 \end{abstract} 
\section{Introduction}
The minimal supersymmetric standard model 
(MSSM), like the standard model itself, predicts a zero 
mass for neutrinos, and this is not compatible 
with the 
recent 
neutrino observations. One of the most promising 
methods to 
attribute a tiny but nonzero mass to neutrinos is the 
seesaw mechanism \cite{seesaw},
which requires three 
extremely 
heavy right-handed neutrinos.  In the MSSM with
right-handed neutrino supermultiplets, the 
leptonic part of the superpotential is 
\be
W=Y_l^{i j} \epsilon_{\alpha \beta}H_1^\alpha   
l_{Ri}^c  L_j^\beta+ Y_\nu^{i j}
\epsilon_{\alpha \beta}H_2^\alpha \nu_{Ri}
L_j^\beta+\frac{1}{2}M_{ij} \nu_{Ri} \nu_{R j},  
\ee
where $L_j^\beta$ is the supermultiplet
corresponding to the 
 doublet $(\nu_{Lj},~ l_{Lj})$. Without loss of 
generality, we can rephase and rotate the 
fields to make the matrices
$Y_l^{ij}$ and $M_{ij}$  real and diagonal:
$Y_l^{ij}={\rm diag}(Y_e, Y_\mu, Y_\tau)$ and 
$M^{ij}={\rm 
diag}(M_1, M_2, M_3)$. In this basis, $Y_\nu$ can 
have off-diagonal and complex elements. Soft
supersymmetry breaking terms of the Lagrangian in 
the context of this 
model can
include
\be -{\cal L}^{soft}_{\tilde{\nu}_R}= (m_0^2)^i_j
(\tilde{\nu}_R^i)^\dagger
\tilde{\nu}_R^j+[\frac{1}{2}
B_\nu 
M^{ij}\tilde{\nu}_R^i\tilde{\nu}_R^j+{\rm 
H.c.}]. 
\label{b}
\ee
The second term in Eq.~(\ref{b}), the ``neutrino 
$B$-term," is a lepton-number-violating 
term \cite{mail} which can cause  profound effects 
 including sneutrino-antisneutrino oscillation 
\cite{yu,hall}. The parameter 
$B_\nu$ is allowed to be 
much larger than the electroweak scale because it is  
 associated  only with $\tilde{\nu}_R$, which is 
an 
electroweak singlet.
It 
has been shown that a
nonzero neutrino $B$-term can create neutrino 
mass 
through 
one-loop diagrams \cite{yu}. 
The upper bound on the neutrino mass  can then be 
translated into 
a
bound on 
$B_\nu$,
\be \label{howard}
B_\nu<10^3 m_{0}.
\ee
If $B_\nu$ is large, some new effects are 
expected 
both in the $e^+e^-$ accelerator experiments 
\cite{yu} 
and in cosmology 
\cite{hall}. In particular,   values of 
$B_\nu$ close to the saturating  bound 
(\ref{howard}) can 
induce observable slepton-antislepton 
oscillation. 
 In this paper, we show that large values of 
$B_\nu$ 
can 
also affect other observables.

It is well-known that nonzero flavor-number-violating slepton
mass terms in the soft Lagrangian ($m_{\alpha 
\beta}^2 
\tilde{L}_\alpha^\dagger \tilde{L}_\beta,  \ \ 
\alpha \ne \beta$) can give 
rise  to rare decays such as ($\mu \to \gamma e$),
($\tau \to \gamma e$), and ($\tau \to \gamma 
\mu$). One way to avoid  flavor changing neutral current (FCNC) effects 
is to choose the 
off-diagonal mass terms to be small.
In fact, theories such as minimal supergravity (mSUGRA) suggest that at 
the grand unified theory (GUT) 
scale, the soft supersymmetry breaking terms are flavor 
blind; that is,  at the GUT scale
 \ba
-{\cal 
L}_{soft} 
&=& 
m_0^2(\tilde{L}_{L\alpha}^\dagger\tilde{L}_{L\alpha}+ 
\tilde{l}_{R\alpha}^\dagger \tilde{l}_{R\alpha}+ 
\tilde{\nu}_{R\alpha}^\dagger 
\tilde{\nu}_{R\alpha}+H_1^\dagger 
H_1+ H_2^\dagger H_2 \label{soft})
+
 \frac{1}{2} m_{1/2}(\tilde{B}^\dagger \tilde{B}+ 
\tilde{W^a}^\dagger \tilde{W^a})
\cr
&+&
(b H_1 H_2 +{\rm H.c.})+
a_0 (Y_l^{ij}\epsilon_{\alpha 
\beta} H_1^\alpha\tilde{l}_{Ri}^\dagger  
\tilde{L}_{Lj}^\beta+ Y_\nu^{ij}\epsilon_{\alpha\beta} 
H_2^\alpha
\tilde{\nu}_{Ri} \tilde{L}_{Lj}^\beta) \cr &+&
\left(\frac{1}{2} B_\nu M_i 
\tilde{\nu}_R^i\tilde{\nu}_R^i+{\rm H. 
c.}\right), 
\ea with universal $m_0^2$, $m_{1/2}$, and $a_0$.

The off-diagonal elements of  the neutrino 
Yukawa coupling radiatively produce 
nonvanishing off-diagonal mass terms for
the left-handed slepton doublet:
 \be \label{past}m_{(1)\alpha 
\beta}^2=-\sum_k {Y_\nu^{k\alpha} 
(Y_\nu^{k\beta})^* \over 16 \pi^2}\left\{ m_0^2\left(3\log 
\left[\frac{\Lambda_{GUT}}{M_k}\right]^2-1\right)+a_0^2 
\log\left[\frac{\Lambda_{GUT}}{M_k}\right]^2\right\}.
\ee             
This effect was first
 noticed and studied in \cite{masiero} and then 
worked out in a series of papers  ({\it e.g.,} see  
\cite{tobe}).
However, the contribution to $m_{\alpha 
\beta}^2$ from   
the neutrino $B$-term
was 
ignored. In section~2, we study this effect and show 
that if $B_\nu$ is large, its contribution will 
dominate over
the effects in 
Eq.~(\ref{past}).

In the MSSM with flavor blind soft supersymmetry breaking 
terms, in addition to  the phases in the Yukawa couplings, 
there are  two 
other  independent CP-violating phases, usually 
chosen to be the phases of the $a_0$ and 
$ \mu$ parameters. These phases can create electric 
dipole moments (EDMs) for charged 
leptons 
and for the 
neutron \cite{amu}. In the presence of the 
neutrino $B$-term, 
there is one 
more 
phase which can also give a contribution to 
the EDM of 
charged leptons. In 
section~3, we show that, 
even if at 
the GUT scale no $A$-term is present ($a_0=0$), 
through one-loop corrections, the 
neutrino $B$-term creates $A$-terms for 
leptons at the 
electroweak scale. This effect could be the 
dominant  term in lepton EDMs.

In section 4, we explore the upper  bounds on 
the imaginary 
and real parts of $B_\nu$ resulting from the upper 
bounds on 
the branching ratios of the rare decays [BR$(l_\alpha 
\to l_\gamma +\gamma)$] and the EDMs of the charged 
leptons.
The main limitation will be the uncertainty in the 
pattern of 
neutrino Yukawa couplings $Y_\nu$.

\section{Effects of the neutrino $B$-term on  
slepton mixing}
It has been shown  that the 
off-diagonal slepton masses ($m_{\alpha \beta}^2 
\tilde{L}_{L\alpha}^\dagger \tilde{L}_{L\beta}$, 
$\alpha \ne \beta$) at the one-loop level  can give rise to 
lepton-number-violating rare decays such as $(\mu 
\to e \gamma)$, $(\tau \to \mu \gamma)$, and
$(\tau \to e \gamma)$ \cite{tobe,hisano}. In the mass insertion 
approximation,
a simplified formula  can be derived 
\cite{casas}:
\be
\label{simple}
{\rm Br}(l_\alpha \to l_\beta+\gamma)\sim {\alpha^3 
\over G_F^2} {\ \ |m_{\alpha \beta}^2|^2 \over 
m_{susy}^8}\tan ^2 \beta.
\ee
The  upper bounds on the  branching ratios of 
the rare decays \cite{pdg} can be interpreted 
as bounds on the off-diagonal elements of 
$|m_{\alpha \beta}^2|$:
\be \label{mue}
|m_{e \mu}^2|< {2 \times 10^{-3} \over \tan 
\beta}({m_{susy} \over 200 \ \ {\rm 
GeV}})^2m_{susy}^2, \ \ \ \ 
|m_{\tau \mu}^2|< {0.4 \over \tan\beta}({m_{susy} 
\over
200 \ \ {\rm GeV}})^2m_{susy}^2 \ \ \ 
\ee
and 
\be \label{etau} |m_{\tau e}^2|< {1  \over 
\tan\beta}({m_{susy}
\over
200 \ \ {\rm GeV}})^2m_{susy}^2.
\ee
The next generation of  experiments \cite{next} 
is expected to  
improve Eq. (\ref{mue}) to
\be \label{ayande}
|m_{e \mu}^2|< {6
\times
10^{-5} \over \tan
\beta}({m_{susy} \over 200 \ \ {\rm
GeV}})^2m_{susy}^2, \ \ \ \ |m_{\tau \mu}^2|< {0.07\over
\tan\beta}({m_{susy}\over
200 \ \ {\rm GeV}})^2m_{susy}^2.\ee

The off-diagonal mass terms for left-handed 
sleptons receive a contribution from the neutrino 
$B$-term through the two diagrams shown in Fig.~1. 
The two lepton number violating vertices on the 
neutrino line are the neutrino $B$-term and the 
standard $\tilde{\nu}_R$ mass term. The neutrino 
$A$-term is also needed.
 The amplitude corresponding to  
diagram~(a) is equal to
\begin{eqnarray} -i{\cal M}=& \sum_k &i Y_{\nu}^{k 
\alpha} 
i(a_0 Y_\nu^{k 
\beta} )^*
\int
{d^4 k \over (2 \pi)^4} {i \over k^2} {i \over
k^2-M_k^2}(-i B_\nu  M_k) {i \over k^2-M_k^2} 
iM_k i
\cr =&\sum_k& {i\over (4 \pi)^2} a_0^* Y_{\nu}^{k 
\alpha} 
(Y_{\nu}^{k   
\beta})^* B_\nu.
\end{eqnarray}
Similarly,   diagram (b) gives
\be
-i {\cal M}=\sum_k {i\over (4 \pi)^2} a_0 Y_{\nu}^{k 
\alpha} 
(Y_{\nu}^{k 
\beta})^* B^*_\nu.
\ee 
The mass correction is given by the sum of the two 
amplitudes:
\be m_{(2)\alpha \beta}^2= -2\sum_k {Y_\nu^{k 
\alpha} 
(Y_\nu^{k 
\beta})^* {\rm Re}[a_0 B_\nu^*]\over (4 \pi)^2}  
,\label{new}
\ee
which has to be added to $m_{(1)\alpha \beta}^2$  
presented in  
Eq.~(\ref{past}). For $ {\rm Re}[a_0 B_\nu^*]\stackrel {>}{\sim}10 m_0^2$,
$m_{(2)\alpha \beta}^2$ exceeds $m_{(1)\alpha \beta}^2$.
Note that the  contribution  we have found does not depend 
on the 
heavy right-handed masses at all. This can be traced 
back to the form
of the  neutrino $B$-term assumed in 
Eq.~(\ref{soft}). 
Had 
we defined this term  as $B_\nu ^2 \
 \tilde{\nu}_R\tilde{\nu}_R$, the result would have been 
proportional to $ {\rm Re}[a_0 (B_\nu^2)^* ] /M_k$.

Up to  factors of $\log (\Lambda_{GUT}/ 
M_k)$,
$m_{(1)\alpha \beta}^2$ and $m_{(2)\alpha \beta}^2$
[see Eqs. (\ref{past},\ref{new})] have the same flavor 
structure. The structure can be different only if the 
masses of right-handed
neutrinos are hierarchical ($M_1\ll M_2 \ll M_3$).
Although the one-loop mass matrix  presented in 
Eq.~(\ref{past}) is enhanced by a factor of $6 \log 
(\Lambda_{GUT} / M_k)\sim  10$,  the neutrino 
$B$-term 
contributions 
given in Eq.~(\ref{new})  dominate if
$B_\nu\sim 10^3 m_0$ as allowed by Eq. (\ref{howard}).

The dependence of $m_{(2)\alpha \beta}^2$ on 
$B_\nu$ involves the combinations $\sum_k 
Y_\nu^{k \alpha} 
(Y_\nu^{k
\beta})^*$. To derive conclusive 
bounds on $B_\nu$, first we have to find some 
lower 
bounds on the   $\sum_k Y_\nu^{k \alpha}
(Y_\nu^{k
\beta})^*$ combinations; however, this information is not 
available at the moment. 
If $Y_\nu$ are so large that
$m_{(1)\alpha \beta}^2$
[see Eq.~(\ref{past})] saturate the bounds
(\ref{mue},\ref{etau}),
$B_\nu$ has to  be smaller than $10 m_0$
[to be
compared with
Eq.~(\ref{howard})].
In section 4, we will discuss this further.

In the discussion above, we have assumed $a_0\sim 
m_0\sim 
m_{susy}$ but it is possible that
$a_0$ is much smaller than $m_0$. In this case,
$m_{(2) \alpha \beta}^2$ given in Eq.~(\ref{new}) will 
not be the dominant effect.
At the one loop level, there is no contribution 
to 
$m^2_{\alpha \beta}$ proportional to $|B_\nu|^2$: it 
can 
be shown that the two 
one-loop
diagrams that are proportional to  $|B_\nu|^2$ 
(depicted in 
Fig.~2) cancel each other
at zero external momentum. However, 
at the two-loop level, there is a contribution proportional to 
$|B_\nu|^2$ which can dominate over $m_{(1)\alpha 
\beta}^2$ 
[Eq.~(\ref{past})] provided that 
$|B_\nu|^2Y_\nu Y_\nu^*/(4\pi)^2>m_0^2$.

\section{Effects of the neutrino $B$-term on 
$A$-terms}
In this section, we show that the neutrino  
$B$-term 
creates an
$A$-term for leptons through one-loop 
diagrams.
We then discuss how this will affect 
the EDMs.
 
As  is discussed in the literature 
\cite{amu}, the phases of  $\mu$ and $a_0$ can 
create electric dipole moments for charged 
leptons as well as for the neutron.
The current bounds on lepton EDMs are  
\be d_e<1.5\times 10^{-27} \ \ e~{\rm cm} 
\ \
{\rm \cite{10}} \ \ \ \ \ d_\mu <7\times 
10^{-19} \ \ e~{\rm cm} \ \ {\rm 
\cite{pdg}}
\ee
and
\be
d_\tau< 3\times 10^{-16}  \ \ e~{\rm cm}\ 
\ 
{\rm \cite{pdg}}.
\ee 
Proposed future experiments are expected to  set 
stronger 
bounds:
\be
d_e<10^{-32}
 \ \ e~{\rm cm}\ \
{\rm \cite{12}} \ \ \ \ 
d_\mu<10^{-24} \ \ (5 \times 10^{-26}) \ \ 
e~{\rm cm}\ 
\ {\rm \cite{14} \ \ (\cite{15})}.
\ee
The  bounds on the electric dipole 
moments of the charged leptons yield strong 
bounds on the imaginary  parts of $\mu$ 
and $a_0$ \cite{isabella}.

The phase of the neutrino $B$-term can provide yet 
another source of CP-violation.
\footnote{
Within the extended MSSM, the Yukawa couplings 
($Y_\nu$) are 
another 
source of CP-violation. 
This effect has been discussed in \cite{us}. } 
When we fixed   the mass matrix $M$ to
be real, the phases of $\tilde{\nu}_R$ were 
fixed; therefore, the phase of $B_\nu$ in 
this convention cannot be 
removed. We expect the imaginary part of 
$B_\nu$
to give contribution to EDMs. 

The parameter $B_\nu$ contributes to the 
$A_l$-term through the diagram shown in Fig.~3.
Adding this correction to the tree level $A_l$
 [see Eq.~(\ref{soft})], we find 
 
\ba
\label{al}
-iA_l^{ji} &=& -ia_0Y_l^{ji}\delta_{ij}+ (i 
Y_l^{jj}) 
(i) i 
(Y_\nu^{kj})^*(i 
Y_\nu^{ki}) \int {i \over k^2}{i \over 
k^2-M_k^2} (-iB_\nu M_k) {-i M_k \over k^2-M_k^2}
 {d^4k \over (2\pi)^4}
\cr
&=&-ia_0Y_l^{ji}\delta_{ij}
-{i \over (4 \pi)^2} Y_l^{jj} 
(Y_\nu^{kj})^*Y_\nu^{ki}B_\nu .
\ea
Similarly, the  neutrino $B$-term contributes to 
the  the $A_\nu$-term through
the diagram  shown in Fig.~4:
\ba \label{anu}
-iA_\nu^{ki} &=&-ia_0Y_\nu^{ki}+ (i Y_\nu^{qi}) 
(i) i
(Y_\nu^{qj})^*(i
Y_\nu^{kj}) \int {i \over k^2}{i \over
k^2-M_q^2} (-iB_\nu M_q) {-i M_q \over k^2-M_q^2}
 {d^4k \over (2\pi)^4}\cr
&=&-ia_0Y_\nu^{ki}
-{i \over (4 \pi)^2} Y_\nu^{qi}
(Y_\nu^{qj})^*Y_\nu^{kj}B_\nu.
\ea

According to
\cite{isabella}, 
for $m_{0}\sim 200$ GeV, the present bound 
($d_e<10^{-27}$ e~cm) implies
\be \label{already} {\rm Im} (A_l^{ee}) 
\langle 
H_1\rangle /(m_e
m_{0})\stackrel {<}{\sim}0.1.
\ee Since the dependence of $d_e$ on
Im($A_l^{ee}$) is linear, if in the future 
the bound $d_e<10^{-32}$~e~cm is obtained, the above bound will be 
improved to
\be \label{5o}{\rm Im}(A_l^{ee}) \langle
H_1\rangle /(m_e
m_{0})\stackrel {<}{\sim}10^{-6}. \ee 
Assuming that $B_\nu$ is the only source of
CP-violation, the bound in Eq. 
(\ref{already}) can be translated into
\be \label{expl}
{\rm Im} [B_\nu] \sum_k Y_l^{ee}
(Y_\nu^{ke})^*Y_\nu^{ke}<15m_0m_e/\langle 
H_1 \rangle ,
\ee
which can be improved by five orders of 
magnitude in the future. 

The present experimental data do not 
lead to any conclusive bounds on the 
values 
of the neutrino Yukawa couplings. In section 
4, we will discuss how 
future observations and developments can improve 
our knowledge on $Y_\nu$.
In principle, $Y_\nu$ can be as large as 
order 1. (In fact, $Y_\nu$ can be 
even larger than 1; however in this case we cannot treat it 
perturbatively.) 
For $Y_\nu^{k e}(Y_\nu^{k e})^*\sim 1$, the present bound 
(\ref{already})  gives Im$[B_\nu]<10m_0$
[we have used Eq.~(\ref{al})
and $m_e=\langle H_1 \rangle Y_l^{ee}$].
Future EDM experiments can make the bound dramatically 
stronger.

Discovery of a lepton EDM could  provide invaluable 
information on $B_\nu$. The neutrino $B$-term gives 
a contribution
to $A_l$; however, it has  
no impact on the $A$-term of quarks. As a result,
Im($B_\nu$) will not affect the EDM of 
the neutron.
On the other hand, Im($a_0$) and Im($\mu$)
give contributions to both  the EDM of 
charged 
leptons and neutron. It is 
possible 
that the contributions of Im($a_0$) and 
Im($\mu$) cancel each other. However, it has been 
shown \cite{falk} that cancellation in the 
electron EDM occurs in the same regions as  
cancellation in the neutron EDM. Therefore, if 
$d_e$ 
turns out to be nonzero while $d_n\ll d_e$,
the effect cannot be attributed to the 
contribution of Im($a_0$) or Im($\mu$). Such a 
situation can be explained with
a nonzero complex $B_\nu$. 

 There is another point 
 that is noteworthy:
EDMs are proportional to $\sum_k| Y_\nu^{k 
\alpha}|^2$, 
and
BR($l_\alpha \to l_\beta \gamma$) are given 
by
 $|\sum_k Y_\nu^{k \alpha}(Y_\nu^{k \beta})^*|^2$, 
which 
 are both 
independent of $M_l$. To the author's
 best knowledge, there is no other 
observable that depends on these combinations.
If $|B_\nu|$ is large, by studying these observables 
we 
can
extract additional information on Yukawa 
couplings which will improve our current understanding 
of the seesaw mechanism and leptogenesis.
 
If the neutrino  $B$-term gives the dominant  
contribution to the electric 
dipole moments, we expect $d_\tau/(m_\tau 
\sum_k |Y_\nu^{k 
\tau}|^2)= d_\mu/(m_\mu \sum_k |Y_\nu^{k \mu }|^2)= 
d_e 
/(m_e \sum_k|Y_\nu^{k e}|^2)$; therefore, if 
$d_e$ is close to its present upper bound, 
$d_e \sim 10^{-27}~e$~cm, we expect 
$d_\mu\sim 10^{-25}~e$~cm which can be 
tested
in  proposed experiments  \cite{15}. 
\section{Bounds on $B_\nu$}
In sections 2 and 3, we have shown that 
large values of $B_\nu$
 can lead to  flavor-violating rare decays 
and EDMs of charged leptons.
However, the dependence of these 
observables on $B_\nu$ is through the unknown 
combination of Yukawa couplings $Y_\nu^{k 
\alpha}(Y_\nu^{k 
\beta})^*$. To derive upper bounds on 
$B_\nu$, we have to find other observables that 
provide lower bounds on these combinations. 
In this section, we  combine various pieces of  
information
on the Yukawa couplings 
(some of them  yet to be obtained) 
to derive a lower bound on  
the factors
$Y_\nu^{k \alpha}(Y_\nu^{k
\beta})^*$.
We will then use the current upper bounds on the 
branching ratios of the rare decays  and the 
values of EDMs to extract upper bounds on $B_\nu$.

Neutrino masses depend on the Yukawa couplings 
through
\be m_{\alpha \beta}^{(\nu)}=\sum_k Y_{\nu}^{k 
\alpha} {1\over M_k}  Y_{\nu}^{k \beta}\langle H_2 
\rangle ^2 .
\ee
All the parameters involved have to be 
evaluated at the electroweak scale. The 
effect of running from the GUT 
scale to the electroweak scale can change 
the details of the neutrino masses and 
mixing \cite{Frigerio:2002in}, however,
the order of  magnitude of the masses 
will not be affected.
Since, here, we want to estimate only the 
order of magnitude of the Yukawa couplings,
we will neglect the running effects.

Currently we  have only bounds on the 
neutrino masses 
\cite{mass}:
$$ \sqrt{\Delta m_{atm}^2}<\sum m_\nu<1~{\rm 
eV},$$
where $\Delta m_{atm}^2=2.5\times 10^{-3} ~ {\rm 
eV}^2$ \cite{atm}.
Future terrestrial and cosmological experiments 
will improve these bounds.
Our knowledge  of the masses of the right-handed 
neutrinos ($M_k$) is 
even less complete than the information on $m_{\alpha 
\beta}^{(\nu)}$. 
If 
leptogenesis is the mechanism behind the 
baryon  
asymmetry of the universe \cite{lepto}, it will be possible 
to derive a lower bound on $M_k$ \cite{uplep}.
For a given value of $m_{\alpha \beta}^{(\nu)}$, 
there is 
at 
least  one $k$ such that
	\be |Y_\nu^{k \alpha}(Y_\nu^{k 
\beta})^*|>\frac{1}{3}\left({m_{\alpha \beta 
}^{(\nu)} 
\over 
0.1 \ \ {\rm eV}}\right){2 \times 10^{-6} 
\over \sin^2 
\beta}\left({M_k \over 6  \times 10^8~{\rm 
GeV}}\right).
\ee

The parameter $m_{ee}^{(\nu)}$ can be extracted 
directly from neutrinoless double decay 
observations. 
If $m_{ee}^{(\nu)}$ is 
relatively large ($m_{ee}^{(\nu)}\sim 0.1 ~{\rm 
eV}$), its effect should be observable in future 
experiments \cite{mee}.
Using Eq.~(\ref{al}),
$d_e<10^{-27}$ e~cm [see 
Eq.~(\ref{already})] yields
$$ {{\rm Im}(B_\nu) \over m_{0}}<3\times 10^{7} 
\left({0.1 ~ 
{\rm  eV} \over 
m_{ee}^{(\nu)}}\right)\left({ 6 \times 
10^8~{\rm GeV} \over 
M_k}\right).$$
In future, if the bound $d_e <10^{-32}$ is 
obtained, this bound will be improved by five 
orders of magnitude [see Eq. (\ref{5o})] 
which 
means it can be 
more restrictive than the bound  in Eq.~(\ref{howard}).

Extracting bounds on $Y_\nu^{k \alpha}(Y_\nu^{k 
\beta})^*$, $\alpha \ne \beta$, will be more 
challenging 
 because, unlike 
$m_{ee}^{(\nu)}$, $m_{\alpha 
\beta}^{(\nu)}$,  
$\alpha \ne \beta$,
cannot be directly measured.
Even if forthcoming experiments find that the 
overall neutrino mass is of order of $0.1$ eV 
(quasidegenerate mass scheme), it 
will be difficult to derive definite lower bounds on
[$m_{\alpha \beta}^{(\nu)}, \ \ 
(\alpha \ne \beta)$]. In the case of  
a quasidegenerate  
mass scheme (the scheme for which the 
absolute values 
of the mass eigenvalues are much larger than  
$\sqrt{\Delta m_{atm}^2}$) with
 zero Majorana 
phases we expect $m_{e\mu}^{(\nu)}, m_{\mu 
\tau}^{(\nu)}, m_{e 
\tau}^{(\nu)}\ll m_{ee}^{(\nu)}, m_{\mu 
\mu}^{(\nu)}, m_{\tau \tau}^{(\nu)}$.
Only in the framework of the
quasidegenerate neutrino mass scheme with at 
least one nonzero 
Majorana phase, it is possible to have large 
off-diagonal neutrino masses, 
$m_{e\mu}^{(\nu)}, m_{\mu   
\tau}^{(\nu)}, m_{e
\tau}^{(\nu)}\gg \sqrt{\Delta m_{atm}^2}$  
(a phase equal to $\pi$ also works).
On the other hand, determining the values of 
Majorana phases is very challenging, if possible at 
all \cite{glashow}.
Nevertheless, let us suppose that in 
the future some 
hypothetical experiment will be able to 
determine $m_{e \mu}^{(\nu)}$. 
Then, assuming that the factors $ Y_\nu^{k 
e}(Y_\nu^{k \mu})^*$ do not cancel each other, the 
present  
bound  on $|m_{e \mu}^2|$ given in Eq.~(\ref{mue}) 
implies
\be {\rm Re}[a_0 B_\nu^*]/m_{0}^2<
{10^5 \over \tan 
\beta} 
\left({m_{0} 
\over 200 \ \ {\rm GeV}}\right)^2  
\left({0.1 \ \ {\rm eV} \over m_{e 
\mu}^{(\nu)}}\right)\left({6 \times 
10^8~{\rm GeV} 
\over M_k}\right). \ee
The future possible bounds [inferred from 
Eq.~(\ref{ayande})]  can 
be  more 
restrictive 
than Eq.~(\ref{howard}): 
\be {\rm Re}[a_0 B_\nu^*]/m_{0}^2<{3\times 10^3 
\over \tan \beta}\left({m_{0}\over 200 \ \ 
{\rm 
GeV}}\right)^2 \left({0.1 \ \ {\rm eV} 
\over m_{e \mu}^{(\nu)}}\right)
\left({6 \times 
10^8~{\rm GeV} \over M_k}\right);\ee
however, as we pointed out earlier, at the moment,
measuring $m_{e \mu}^{(\nu)}$ seems to be 
impossible.

\section{Concluding remarks}
We 
have studied the effects of the neutrino $B$-term 
on
the slepton 
mixing and  EDMs of charged leptons in  
the framework of seesaw model 
embedded in the
MSSM with universal soft supersymmetry breaking 
terms.

If $B_\nu>10 m_0\sim 10 a_0$ but $a_0 > B_\nu Y_\nu 
(Y_\nu)^*/(4 \pi)^2$,
the dominant flavor-violating slepton masses are 
given by Eq.~(\ref{new}) rather than 
Eq.~(\ref{past}).
For values of $B_\nu$ satisfying $B_\nu Y_\nu 
(Y_\nu)^*/(4\pi)^2>a_0$ and $|B_\nu|^2 
Y_\nu(Y_\nu)^*/(4\pi)^2>m_0^2$, 
the two-loop contribution proportional to $|B_\nu|^2$ 
can be 
dominant.
The bounds on the Yukawa couplings and 
neutrino 
$B$-term which 
have 
been discussed in the literature allow quite large 
contributions to the slepton masses, 
violating  the upper bounds from rare 
flavor-violating decays.
However, since there is no direct lower bound 
on the combinations of neutrino Yukawa couplings 
appearing in the formulations, it is not possible to 
derive any  upper bound on the  
Re[$a_0B_\nu^*$].

The parameter $B_\nu$ can be considered as 
another source 
for CP-violation and therefore EDMs. In fact, we 
have shown
that the neutrino $B$-term directly creates  
$A$-terms both 
for 
neutrinos and charged leptons--but not 
quarks--even 
if $a_0=0$ at the GUT scale. 
The imaginary part of $A_l$ gives a contribution to 
the 
EDMs of charged leptons.
If  the  $B_\nu$ effect  is
dominant, we expect 
$d_\tau/(m_\tau 
\sum_k |Y_\nu^{k
\tau}|^2)= d_\mu/(m_\mu \sum_k |Y_\nu^{k \mu 
}|^2)=
d_e
/(m_e \sum_k|Y_\nu^{k e}|^2)$; therefore, if
$d_e$ is close to its present upper bound we expect
that the proposed experiments \cite{15} 
will be able to 
measure the value of $d_\mu$.
The discovery of nonzero 
$d_e$ and $d_\mu$ while $d_n\ll 
d_e$ can be explained  
 with large
Im($B_\nu$). In this case, $d_e 
\propto \sum_k Y_\nu^{ke}(Y_\nu^{k e})^* {\rm Im} 
(B_\nu).$
If  Im$(B_\nu)$ is determined through some 
other 
observation, information on $d_e$ and Im$(B_\nu)$ 
combined with 
$m_{ee}^{(\nu)}=\sum_k \langle H_2 
\rangle ^2(Y_\nu^{k e})^2/M_k$ 
(extracted from 
neutrinoless double beta decay searches) can provide 
us with information on the values of $M_k$, shedding 
light on the origins of neutrino masses and 
on 
leptogenesis.
\section*{ Acknowledgment}
I would like to thank Y. Grossman, M. 
Peskin, and A. Yu. Smirnov
for useful discussions and encouragement. I am also 
grateful to M. M. Sheikh-Jabbari for careful reading 
of the manuscript.

 
\newpage

\SetWidth{1}
\vspace{-10pt} \hfill \\
\SetScale{1}
\begin{picture}(80,30)(0,50)
\ArrowLine(10,0)(50,0)
\Text(30,10)[]{$\tilde{L}_\alpha$}
\ArrowArc(100,0)(50,0,180)
\Text(100,60)[]{$H_2$}
\ArrowArcn(100,0)(50,240,180)
\Text(45,-35)[]{$F_\nu^k$}
\Text(75,-43)[]{{\bf $\otimes$}}
\Text(100,-60)[]{$\tilde{\nu}_R^k$}
\ArrowArc(100,0)(50,240,300)
\Text(155,-35)[]{$\tilde{\nu}_R^k$}
\Text(125,-43)[]{$\otimes$}
\ArrowArcn(100,0)(50,360,300)
\Vertex(150,0){3}
\ArrowLine(150,0)(190,0)  
\Text(170,10)[]{$\tilde{L}_\beta$}
\Text(90,-80)[]{(a)}
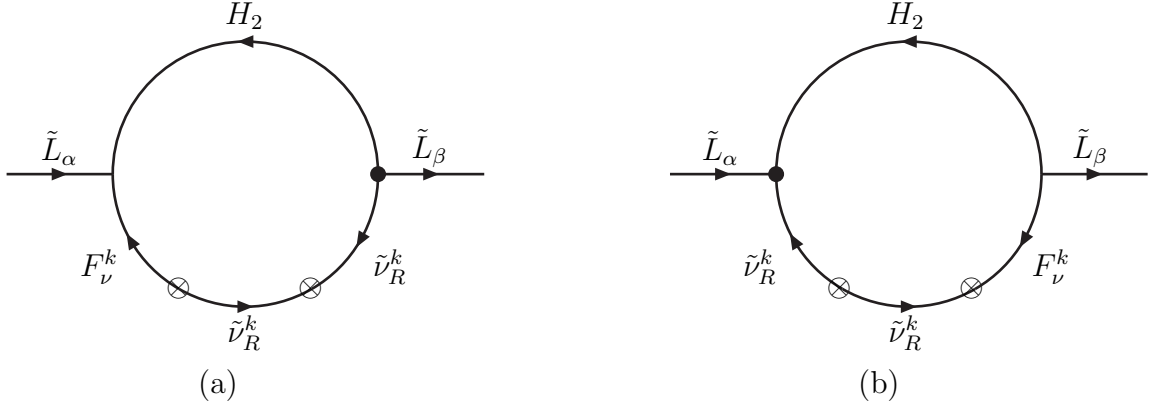
\Text(10, -100)[l] {Figure 1: Diagrams contributing to 
 slepton masses. $F_\nu^k$ represents the auxiliary 
field}
\Text(10, -115)[l]{associated with 
$\tilde{\nu}_R^k$. The $A_\nu$ vertices are marked 
with black circles.} \ArrowLine(260,0)(300,0)
\Text(280,10)[]{$\tilde{L}_\alpha$}
\ArrowArc(350,0)(50,0,180)
\Text(350,60)[]{$H_2$}
\ArrowArcn(350,0)(50,240,180)
\Text(295,-35)[]{$\tilde{\nu}_R^k$}
\Text(325,-43)[]{$\otimes$}   
\Text(350,-60)[]{$\tilde{\nu}_R^k$}
\ArrowArc(350,0)(50,240,300)
\Text(405,-35)[]{$F_\nu^k$}
\Text(375,-43)[]{$\otimes$}
\ArrowArcn(350,0)(50,360,300) 
\ArrowLine(400,0)(440,0)
\Text(420,10)[]{$\tilde{L}_\beta$}
\Text(340,-80)[]{(b)}
\Vertex(300,0){3}

\end{picture}\hspace{3pt}
\vspace{200pt} \hfill \\
\SetScale{1}
\begin{picture}(80,30)(0,50)
\ArrowLine(10,0)(50,0)
\Text(30,10)[]{$\tilde{L}_\alpha$}
\ArrowArc(100,0)(50,0,180)
\Text(100,70)[]{$F_{H_2}$}
\ArrowArcn(100,0)(50,240,180)
\Text(45,-35)[]{$\tilde{\nu}_R^k$}
\Text(75,-43)[]{$\otimes$}
\Text(100,-60)[]{$\tilde{\nu}_R^k$}
\ArrowArc(100,0)(50,240,300)
\Text(155,-35)[]{$\tilde{\nu}_R^k$}
\Text(125,-43)[]{$\otimes$}
\ArrowArcn(100,0)(50,360,300)
\ArrowLine(150,0)(190,0)  
\Text(170,10)[]{$\tilde{L}_\beta$}
\Text(90,-90)[]{(a)} 
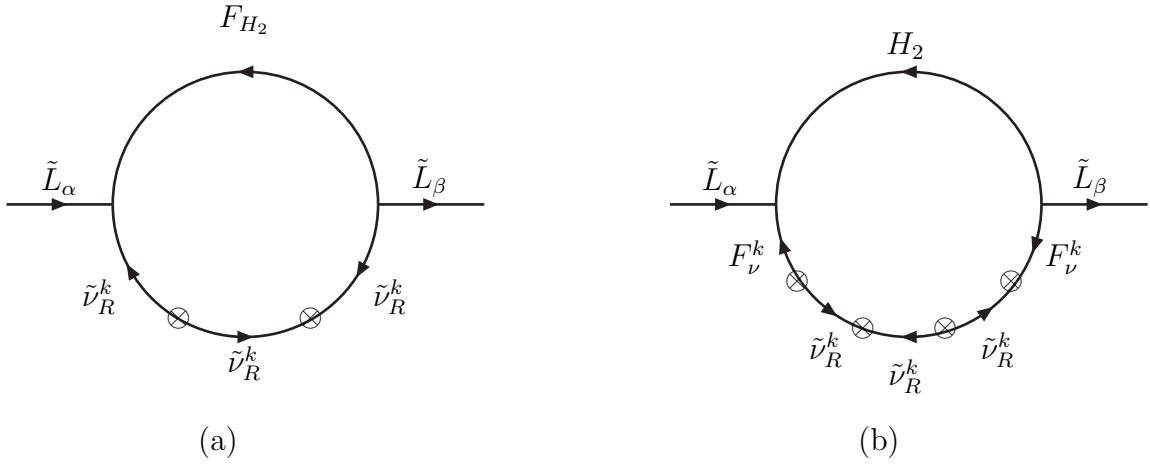
\Text(10, -110)[l]{Figure 2: Diagrams  proportional 
to
$|B|^2$ contributing to slepton masses. $F_{H_2}$ 
represents}
\Text(10, -125)[l]{the auxiliary 
field associated with $H_2$.}

\ArrowLine(260,0)(300,0)   
\Text(280,10)[]{$\tilde{L}_\alpha$}
\ArrowArc(350,0)(50,0,180) 
\Text(350,60)[]{$H_2$}
\ArrowLine(400,0)(440,0)  
\Text(420,10)[]{$\tilde{L}_\beta$}
\Text(340,-90)[]{(b)}
\ArrowArcn(350,0)(50,216,180)
\Text(309,-29)[]{$\otimes$}
\ArrowArc(350,0)(50,216,252)
\Text(334,-47)[]{$\otimes$}
\ArrowArcn(350,0)(50,288,252) 
\Text(365,-47)[]{$\otimes$}
\ArrowArc(350,0)(50,288,324)
\Text(390,-29)[]{$\otimes$}
\ArrowArcn(350,0)(50,360,324) 
\Text(410,-20)[]{$F_\nu^k$}
\Text(290,-20)[]{$F_\nu^k$}  
\Text(320,-55)[]{$\tilde{\nu}_R^k$}
\Text(385,-55)[]{$\tilde{\nu}_R^k$}
\Text(350,-65)[]{$\tilde{\nu}_R^k$}
\end{picture}
\newpage

\vspace{200pt} \hfill \\   
\SetScale{1}
\begin{picture}(80,30)(0,50)
\ArrowLine(160,0)(200,0)     
\Text(180,10)[]{$\tilde{L}_i$}
\ArrowArc(250,0)(50,0,180)
\Text(250,60)[]{$H_2$}
\ArrowArcn(250,0)(50,240,180)
\Text(195,-35)[]{$F_\nu^k$}
\Text(225,-43)[]{$\otimes$}
\Text(250,-60)[]{$\tilde{\nu}_R^k$}
\ArrowArc(250,0)(50,240,300)
\Text(305,-35)[]{$\tilde{\nu}_R^k$}
\Text(275,-43)[]{$\otimes$}  
\ArrowArcn(250,0)(50,360,300) 
\ArrowLine(300,0)(340,0)  
\Text(320,10)[]{$F_L^j$}
\Text(240,-90)[]{(a)}
\Text(10, -110)[l] {Figure 3: Diagram
 contributing to $A_l$. $F_\nu^k$
 and $F_L^j$
represent the auxiliary
fields associated}
\Text(10, -125)[l]{with $\tilde{\nu}_R^k$ and
$\tilde{L}_j$, respectively.}
\ArrowLine(360,50)(340,0)
\Text(340,30)[]{$\tilde{l^c}_{Rj}$}
\ArrowLine(360,-50)(340,0)
\Text(340,-30)[]{$H_1$}
\end{picture}


\vspace{200pt} \hfill \\
\SetScale{1}\begin{picture}(80,30)(0,50)

\ArrowLine(160,0)(200,0)  
\Text(180,10)[]{$\tilde{L}_i$}
\ArrowArc(250,0)(50,0,180)
\Text(250,60)[]{$H_2$}
\ArrowArcn(250,0)(50,240,180)
\Text(195,-35)[]{$F_\nu^q$}
\Text(225,-43)[]{$\otimes$}
\Text(250,-60)[]{$\tilde{\nu}_R^q$}
\ArrowArc(250,0)(50,240,300)
\Text(305,-35)[]{$\tilde{\nu}_R^q$}
\Text(275,-43)[]{$\otimes$}
\ArrowArcn(250,0)(50,360,300) 
\ArrowLine(300,0)(340,0)  
\Text(320,10)[]{$F_L^j$}
\Text(240,-90)[]{(a)}
\Text(10, -110)[l]{Figure 4: Diagram
 contributing to $A_\nu$. $F_\nu^q$ and $F_L^j$
represent the auxiliary
fields associated}
\Text(10, -125)[l]{with $\tilde{\nu}_R^q$ and 
$\tilde{L}_j$, respectively.} 
\ArrowLine(360,50)(340,0)
\Text(340,25)[]{$\tilde{\nu}_{R}^k$}
\ArrowLine(360,-50)(340,0)
\Text(340,-30)[]{$H_2$}

\end{picture}

\end{document}